\documentclass{elsarticle}

\usepackage{pifont, fleqn, geometry}

\usepackage{graphicx, natbib, amssymb}
\usepackage[pdftex,colorlinks,citecolor=blue,bookmarks]{hyperref}
\usepackage[fleqn]{amsmath}
\usepackage{color}
\usepackage{hypernat}

\usepackage{amssymb}
\usepackage{dcolumn}
\usepackage{braket}
\usepackage{bm}
\usepackage{color}
\usepackage{multirow}
\usepackage{threeparttable}
\usepackage{booktabs, dcolumn}
\usepackage{siunitx}
\usepackage{comment}
\usepackage{memhfixc}
\usepackage{graphicx}
\graphicspath{ {figures/} }
\usepackage{gensymb}
\usepackage[mathscr]{euscript}

\usepackage[compat=1.1.0]{tikz-feynman}
\usepackage{tikz}
\usetikzlibrary{shapes,arrows,positioning,automata,backgrounds,calc,er,patterns}
\usepackage{bm}

\makeatother

\usepackage[all]{hypcap}
\usepackage{float}
\usepackage{cleveref}
\usepackage{url}
\usepackage{ulem}

\begin{document}

\begin{frontmatter}

\title{Theoretical study on $\eta'\rightarrow \pi^+\pi^-\pi^{+(0)}\pi^{-(0)} $}


\author[]{Ehsan Jafari\corref{mycorrespondingauthor}}
\cortext[mycorrespondingauthor]{Corresponding author}
\ead{ehsan.jafari80@gmail.com}

\author{Bing An Li}
\ead{bing.li@uky.edu}

\address{Department of Physics and Astronomy, University of  Kentucky, Lexington, KY 40506-0055, USA}

\begin{abstract}

  The $\eta'$ meson is associated with the U(1) anomaly. In this paper, a
  successful effective chiral theory of mesons has been applied to study the anomalous decay
  of $\eta'\rightarrow \pi^+\pi^-\pi^{+(0)}\pi^{-(0)}$. In this study, the contributions of triangle
  and  box anomalies are calculated. It is shown that the contribution of box diagrams is important in this process. We predict
  branching ratios of $Br(\eta' \rightarrow \pi^+ \pi^- \pi^+ \pi^-)={1 \over 2}Br(\eta' \rightarrow \pi^+
  \pi^- \pi^0 \pi^0) = (8.3 \pm 1.2) \times 10^{-5}$, which is in good agreement with BESIII measurement.

\end{abstract}


\end{frontmatter}

\section{Introduction}
\label{Intro}

Anomaly is a fantastic phenomenon of quantum field theory. There are Adler-Bell-Jakiew triangle anomaly~\cite{1,2}
and Chanowitz box anomaly, too \cite{3,4}. On the other hand, the anomalies are described by
anomalous Lagrangian:
Wess-Zumino-Witten~\cite{5,6},
\"{O}. Kaymakcalan, S. Rajeev and J. Schechter~\cite{7}, K. C. Chou, H. Y. Guo, K. Wu, and X. C. Song~\cite{8}.
It is well known that the $\eta'$ meson is associated with the U(1)
anomaly~\cite{9, 10}.
Of course, besides anomalous meson processes there are normal meson processes, for  example,
$\rho\rightarrow \pi\pi$, $\eta' \rightarrow \eta \pi \pi,$ etc. Chiral Perturbation Theory (ChPT) is used to describe those
meson processes with normal parity~\cite{11,12,13,14}.

In the CLEO experiment~\cite{15} upper limits are set on the branching ratios for $\eta' \rightarrow \pi^+\pi^-\pi^{+(0)}\pi^{-(0)}$
and in the BESIII experiment~\cite{16}, the decay modes are observed and branching ratios are determined to be $Br(\eta'\rightarrow
\pi^+ \pi^- \pi^+ \pi^-)=[8.53\pm0.69(stat.)\pm0.64(syst.)]\times 10^{-5}$ and $Br(\eta' \rightarrow \pi^+ \pi^- \pi^0
\pi^0)=[1.82\pm0.35 (stat.)\pm0.18(syst.)]\times 10^{-4}$.

Based on a combination of ChPT and vector meson dominance (VMD), Guo, Kubis and Wirzba performed a theoretical study of the
$\eta'$ decay to charged pions~\cite{17}. In the mentioned study the decay amplitudes are dominated by the triangle anomaly and predicted results for
the branching fractions of $Br(\eta' \rightarrow \pi^+\pi^- \pi^+\pi^-)= (1.0 \pm 0.3) \times 10^{-
4}$ and $Br(\eta' \rightarrow \pi^+ \pi^- \pi^0 \pi^0)= (2.4 \pm 0.7) \times 10^{-4}$ are in agreement with the experiment.

In Refs.~\cite{18,19} an effective chiral theory of pseudoscalar, vector and axial-vector mesons has been proposed.
This effective theory has been successfully applied to study different meson processes~\cite{18, 19, 20, 21, 22, 23, 24, 25, 26, 
27, 28, 29, 30}. For example,  meson decays with normal parity like $a_1 \to \rho \pi$, $\phi \to K^+K^-$, $\eta' \to \eta\pi^+
\pi^-$, $k^\star(892) \to K\pi$, $K_1(1400) \to K^\star(892)\pi$, etc., and anomalous decays $\pi^0 \rightarrow \gamma \gamma$, $
\eta^{(')} \rightarrow \gamma \gamma$, $\eta' \rightarrow \rho \gamma$, $\phi \rightarrow \eta \gamma$, $\omega\rightarrow \pi\pi
\pi$, etc., have been evaluated. We briefly review this theory in the next section. In this paper, we apply this effective chiral 
theory to study anomalous processes $\eta' \rightarrow \pi^+\pi^-\pi^{+(0)}\pi^{-(0)}$. Based on our results, the contribution of 
box anomaly is important to the  decay amplitudes. Our theoretical predictions are in good agreement with the experiment.

This work is organized as follows: In section 2, we briefly review the effective chiral theory of mesons which has been applied
in this paper. In section 3, we calculate $\eta' \rightarrow \pi^+\pi^-\pi^+\pi^-$ branching ratio by evaluating
triangle and box anomalies, and with the use of isospin relation we connect this branching ratio to $\eta' \rightarrow \pi^+\pi^-\pi^0\pi^0$.  We summarize our results in section 4.


\section{Review of the Effective Chiral Theory}
\label{Sec.2}

It is well known that the current algebra is very successful in the study of hadron physics.
Based on current algebra and large $N_C$ expansion of QCD, the Lagrangian of $U(3)_L \times U(3)_R$ chiral field theory of quarks
and mesons ($0^{-+}$, $1^{--}$ and $1^{++}$) has been constructed~\cite{18,19}

\begin{equation}\label{eq.1}
\begin{split}
{\cal L} =& \bar{\psi}\big\{i\gamma\cdot\partial + \gamma\cdot v + eQ\gamma\cdot A+ \gamma\cdot a \gamma_5- m u \big\}\psi
 +{1\over2} m^2_0 (\rho^\mu_i\rho_{\mu i} + \omega^\mu\omega_\mu + a^\mu_i a_{\mu i} + f^\mu f_\mu )\\
& +{1\over 2} m^2_0 ( K^{* a}_\mu \bar{K}^{*a\mu} + K^{\mu}_1 K_{1\mu})
+m^2_0 (\phi^\mu \phi_\mu + f^\mu_s f_{s\mu})\quad\quad\quad\quad\quad\quad\quad\quad\quad\quad\quad\quad \:
\end{split}
\end{equation}

with

\begin{equation} \label{eq.2}
\begin{split}
&u = exp^{\{i \gamma_5 (\pi + K + \eta_8 + \eta_0)\}}\\
&a_\mu = \tau_i a^i_\mu + \lambda_a K^a_{1\mu} + ({2\over3} + {1\over\sqrt{3}}\lambda_8)f_\mu + ({1\over3} - {1\over\sqrt{3}}\lambda_8) f_{s\mu}\\
&v_\mu = \tau_i \rho^i_\mu+ \lambda_a K^{* a}_\mu + ({2\over3} + {1\over\sqrt{3}}\lambda_8) \omega_\mu +({1\over3}-{1\over\sqrt{3}}\lambda_8)\phi_\mu
\end{split}
\end{equation}

where  $i = 1,2,3$ and $a= 4,5,6,7$. The $\psi$ in Eq.~(\ref{eq.1}) is $u,d,s$ quark fields. The scheme of nonlinear $\sigma-$model 
is used to introduce
pseudoscalar mesons into Eq.~(\ref{eq.1}) and parameter $m$ is originated in quark condensation and it leads to the dynamical 
chiral symmetry breaking. In this Lagrangian (Eq.~(\ref{eq.1})) meson fields are coupled to the corresponding quark field 
bilinears.
The $\eta_8$ and $\eta_0$ are octet and singlet, respectively. By assuming the mixing angle $\theta=-20$, $\eta$ and $\eta'$ 
fields are defined as:

\begin{eqnarray}\label{eq.3}
\eta = cos\theta\; \eta_8 - sin\theta\; \eta_0 \nonumber\\
\eta' =sin\theta\; \eta_8 + cos\theta\; \eta_0
\end{eqnarray}

Mesons are bound state solutions of QCD and are not independent degrees of freedom. Thus, in Eq.~(\ref{eq.1})
there are no kinetic terms for meson fields. The kinetic terms of the meson fields are generated from quark loops. This theory is an effective
theory, therefore, a cut-off is necessary to be introduced~\cite{31,32}. In the chiral limit $m_q \rightarrow 0$, the cut-off $\Lambda$ is defined~\cite{18}

\begin{equation}\label{eq.4}
  {N_c\over (4\pi)^2} \Big\{ln(1+{\Lambda^2 \over m^2})+{1 \over 1+ {\Lambda^2 \over m^2}} -1 \Big\}= {1 \over 16} {F^2 \over m^2}
\end{equation}

By normalizing the kinetic terms of pion, $\eta'$ and $\rho$ fields, physical meson fields are defined~\cite{18,19}

\begin{equation}\label{eq.5}
{2\over f_{\pi}} \pi\rightarrow\pi_{physical}\;,\;\;\:\:
{2\sqrt{2}\over f_{\pi}} \eta'\rightarrow\eta'_{physical} \;,\;\;\:\: {1\over g} \rho \rightarrow \rho_{physical}
\end{equation}

$f_\pi$ is the pion decay constant and $g$ is a universal coupling constant which  are defined as

\begin{equation}\label{eq.6}
f_\pi= F^2\big( 1-{2c\over g}\big),\quad\quad g^2={F^2 \over 6m^2}, \quad\quad
c={f_\pi^2 \over 2gm_\rho^2}
\end{equation}

$f_{\pi}$ and $g$ are two inputs and $f_\pi = 186 MeV$  and $g=0.395$  are taken. Thus, the cut-off is determined to be $\Lambda \sim 1.8$ GeV.
All the masses of mesons are below the cut-off and the theory is self-consistent.
The input values of $f_\pi$ and $g$ are chosen such that the theory fits the experimental data for different meson processes~\cite{20,21,22,23,24,25,26,27,28,29,30}.

As shown in Refs.~\cite{18,19} the VMD is a natural result of this meson theory instead of an input. According to Sakurai~\cite{33}, the VMD
is revealed from a Lagrangian in which photon and vector mesons are coupled to quarks symmetrically. These symmetries are
shown in the  Lagrangian (\ref{eq.1}), in which the photon field is added. At the fourth order in covariant derivative,  the $\rho-\gamma$ vertex is derived from the quark vertex of  photon and $\rho$ meson~\cite{18}

\begin{eqnarray}\label{eq.7}
{\cal L} = - {1\over 4}e g (\partial _\mu A_\nu -\partial _\nu A_\mu )(\partial _\mu \rho_\nu -\partial _\nu \rho_\mu),
\end{eqnarray}

where $A$ is the photon field. By using Eq.~(\ref{eq.7}) the $\Gamma^{th}(\rho \to e^+e^-)= 6.75\; KeV$ is predicted. The experimental
value, as quoted by particle data group (PDG), is $\Gamma^{exp}(\rho \to e^+e^-)= 7.04(6)\; KeV$~\cite{34}.  There are similar terms for $\omega-
\gamma\;\text{and}\ \phi-\gamma$ vertex~\cite{18,19}. By employing the VMD of $\rho-\gamma, \;\omega-\gamma,$ and  $\phi-\gamma$,
 the pion form factor, the form factors of the charged and the neutral kaons
are obtained~\cite{27,30}.


In this theory the quark loop is always of order $N_C$. The meson loops are at higher order in $N_c$ expansion and  $f_\pi$ and 
$g$ are both of  order $\sqrt{N_C}$~\cite{18,19}. The meson physics studied are at the leading order of $N_C$ expansion.  In the chiral limit $m_q \rightarrow 0$, the theory is explicitly chiral symmetric. In this limit, $f_\pi$ and $g$ are two parameters.

Many anomalous processes of mesons  have been studied by using this theory. The ABJ anomaly $\pi^0\rightarrow \gamma
\gamma$~\cite{1,2} is obtained from both the vertices $\pi\omega\rho$ (WZW anomaly~\cite{5,6}) and VMD $\rho-\gamma$ and $\omega-
\gamma$~\cite{18}.
Other meson anomalies;
 $\eta^{(')} \rightarrow \gamma\gamma, \; \eta' \rightarrow \omega\gamma, \; \rho \rightarrow \eta \gamma,
\; \omega \rightarrow \eta \gamma,$ etc., have been studied in Refs.~\cite{18,19}. Theoretical results are in good agreement with 
experimental data.

In this effective chiral theory~\cite{18, 19} meson resonances are involved. As pointed out in Refs.~\cite{11,35,36}, the coupling 
constants of effective chiral
Lagrangian for strong interactions are essentially saturated by meson resonance exchange. In this regard, the anomalous decay $\omega\rightarrow 3
\pi$ is very interesting. In this decay $\omega\rightarrow \rho + \pi,\;\rho\rightarrow
\pi\pi$ and direct $\omega\rightarrow 3 \pi$ are involved. The amplitude derived in Ref.~\cite{18} is the same as the one derived 
by
\"{O}. Kaymakcalan, S. Rajeev, and J. Schechter~\cite{4}. The vertex $\omega\rightarrow \rho + \pi$ is from the
triangle diagram of quarks and the direct vertex $\omega\rightarrow 3\pi$ is from the box diagram. Both are in low
energies. The vector meson resonance is not involved in the box diagram and as it is shown, the contribution of the box diagram is 
small. To be more precise, the contribution of box anomaly to the branching ratio $Br(\omega \to 3\pi)$ is only 5\% ~\cite{18}.

At low energies ($E < m_\rho $) the theory (Eq.~(\ref{eq.1})) goes back to ChPT and the 10 coefficients of ChPT are determined~
\cite{24,29}.
This theory has already been applied to study various aspects of meson physics~\cite{18,19,20,21,22,23,24,25,26,27,28,29,30}. 
The theory agrees with the experimental data well.


\section{ Calculation of $\eta'\rightarrow \pi^+\pi^-\pi^+\pi^-$ and $\eta'\rightarrow \pi^+\pi^-\pi^0\pi^0$}

In this paper the effective chiral theory of mesons~\cite{18,19} is applied to study the anomalous decays $\eta'\rightarrow \pi^+\pi^- \pi^{+(0)}\pi^{-(0)} $. Before doing
this study it is interesting to mention that the decay width of $\eta'\rightarrow \eta \pi\pi$ with normal parity is computed in
Ref.~\cite{19}.

\begin{figure}[ht]\label{fig1}
\centering
\includegraphics[scale=0.5]{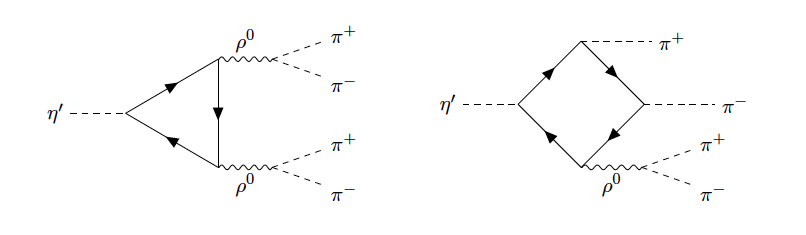}
\caption{Feynman diagrams for triangle and box anomalies contributing to $\eta' \to \pi^+ \pi^- \pi^+ \pi^-.$ Feynman diagrams were generated using TikZ-Feynman~\cite{38}}
\end{figure}

In this process there are two kinds of anomaly: 1) triangle anomaly~\cite{1,2} of two body decays $\eta'\rightarrow \rho \rho,
\rho\rightarrow\pi \pi$; 2) box anomaly~\cite{3,4} of three body decays $\eta'\rightarrow \rho\pi\pi,\rho\rightarrow\pi\pi$. The vertex
$\eta'\rightarrow \rho \rho$ comes from the triangle diagram of quarks and the vertex $\eta'\rightarrow \rho \pi\pi$ comes from
the box diagram, shown in Fig.(\ref{fig1}). The box anomaly proposed by Chanowitz has been applied to study the three body decay of $\eta'\rightarrow\pi^+\pi^-\gamma$ in Ref.~\cite{37}.

In the theory~\cite{18,19}, the pion and $\eta'$ fields have two sources:
one from the term $u$ of the Lagrangian (Eq.~(\ref{eq.1})), the other from the shift caused by the mixing between the axial-vector and
pseudoscalar fields

\begin{equation}\label{eq.8}
\begin{split}
&a^i_\mu \rightarrow {1\over g}(1 - {1\over 2\pi^2 g^2})^{- {1\over 2}} a^i_\mu - {c\over g}{2\over f_\pi} \partial_\mu \pi^i,\;\;\;\\
&f_{\mu} \rightarrow {1\over g}(1 - {1\over 2\pi^2 g^2})^{- {1\over 2}} f_{\mu} - {c\over g}{2\sqrt{2}\over f_{\pi}} \partial_\mu \eta'.
\end{split}
\end{equation}

Eqs.~(\ref{eq.8}) are the result of mixing  $a_\mu$ and $\partial_\mu \pi^i$, and $f_\mu$ and $\partial_\mu\eta'$, which are
generated by corresponding quark loop diagrams. The constant ${1\over g} (1- {1 \over {2\pi^2 g^2}})^{-{1 \over 2}}$ is the
renormalization constant of the $a^i_\mu$ and $f_{\mu}$ fields.

In Ref.~\cite{19} the triangle anomaly of the $\eta'$ is expressed as

\begin{equation}\label{eq.9}
\begin{split}
{\cal L}_{\eta'vv} =& \frac{N_C}{(4\pi)^2}{4\over g^2}{2\sqrt{2}\over f_\pi}
\varepsilon^{\mu\nu\alpha\beta}\eta' \Big\{(\sqrt{2\over3} cos\theta +
{1\over \sqrt{3}}sin\theta) (\partial_\mu\rho^i_\nu\partial_\alpha\rho^i_\beta + \partial_\mu\omega_\nu\partial_\alpha\omega_
\beta)+
 \\
&(\sqrt{2\over3} cos\theta -{2\over \sqrt{3}}sin\theta) \partial_\mu\phi_\nu\partial_\alpha\phi_\beta\Big\},
\end{split}
\end{equation}

where the $\theta$ is the mixing angle.

The $\eta'\rightarrow \gamma\gamma; \;\rho\gamma; \;\omega\gamma$ anomalous decay modes have been studied by using
this Lagrangian (\ref{eq.9}).
The theory agrees with experimental data without a new parameter~\cite{19}.

In this paper, the triangle anomaly $\eta'\rightarrow \rho^0 \rho^0
\rightarrow \pi^+ \pi^- \pi^+ \pi^-$ is studied by first applying Lagrangian (\ref{eq.9}).
The triangle amplitude is expressed as Eq.~(\ref{eq.10})

\begin{eqnarray}\label{eq.10}
T(1) = \frac{N_C}{(4\pi)^2} {2\sqrt{2}\over f_\pi}{32\over g^2}(\sqrt{2\over 3} cos\theta +{1\over\sqrt{3}}sin\theta) m_{\eta'}
\epsilon_{ijk} k_{1i} k_{2j} k_{3k}\nonumber\\
\Big\{\frac{f_{\rho\pi\pi}(q^2_1)}{q^2_1 - m^2_\rho +i\sqrt{q^2_1} \Gamma_\rho (q^2_1)} \frac{f_{\rho\pi\pi}(q^2_2)}
{q^2_2 - m^2_\rho +i\sqrt{q^2_2} \Gamma_\rho (q^2_2)} \nonumber\\
- \frac{f_{\rho\pi\pi}(q^2_3)}{q^2_3 - m^2_\rho +i\sqrt{q^2_3} \Gamma_\rho (q^2_3)}
\frac{f_{\rho\pi\pi}(q^2_4)}{q^2_4 - m^2_\rho +i\sqrt{q^2_4} \Gamma_\rho (q^2_4)}\Big\},
\end{eqnarray}

where \(q_1 = k_1 +k_2 ; \ q_2 = k_3 +k_4 ;\ q_3 = k_1 +k_4 ;\ q_4 = k_2 +k_3 \) and $k_i,\;\;i = 1, 2, 3, 4$ are the momentum of
the four pions, respectively, and

\begin{equation}\label{eq.11}
\begin{split}
&f_{\rho\pi\pi} = {2\over g}\Big\{1+\frac{q^2}{2\pi^2 f^2_\pi} \big[(1 - {2c\over g})^2 - 4 \pi^2 c^2 \big]\Big\},\\
&\Gamma (\rho\rightarrow\pi\pi) = \frac{f^2_{\rho\pi\pi}(q^2)}{48\pi}\sqrt{q^2}\big( 1 - \frac{4 m^2_\pi}{q^2}\big)^{3\over2},
\end{split}
\end{equation}

where $q$ is the momentum of the $\rho$ meson. Taking \(q^2 = m^2_\rho\), and using $m_\rho=775.26\; MeV$  as quoted by PDG~\cite{34}, $\Gamma_\rho = 146\; MeV$ is obtained.

Besides the two body decay channel, the decay $\eta'\rightarrow \pi^+\pi^-\pi^+\pi^-$ have the channel of three body decay $\eta'\rightarrow \rho
\pi\pi$. As we mentioned previously, in the study $\omega\rightarrow 3\pi$ both the triangle,
$\omega\rightarrow \rho\pi,\;\rho\rightarrow 2\pi$, and the box diagram $\omega\rightarrow 3\pi$ are
included and the contribution of the box diagram is very small~\cite{18}.
But, for the $\eta'\rightarrow \rho +  2\pi,\;\rho\rightarrow2\pi$,
the $\rho$-resonance  boosts the contribution of box diagram~\cite{11,35,36}. Thus,
the box anomaly~\cite{3,4} $\eta'\rightarrow \rho^0  \pi^+ \pi^-,\;\rho^0\rightarrow \pi^+ \pi^-$ must be included.

From Eqs.~(\ref{eq.1},\ref{eq.8}) we can see that $\pi$ and $\eta'$ are coupled to pseudoscalar and axial-vector currents and  $\rho^0$ is coupled to  vector current. Thus, by permuting the final states in the box diagram  $\eta'\rightarrow \rho^0 \pi^+ \pi^-$, we can see that there are 48 box diagrams.  The effective Lagrangian  of all box diagrams is obtained as

\begin{eqnarray}\label{eq.12}
{\cal L}_{\eta'\rho\pi\pi} = {1\over g} \frac{8\sqrt{2}}{f^3_\pi}\frac{N_C}{3\pi^2}
({1\over2}-{2c\over g}+ {c^2\over g^2}) \varepsilon^{\mu\nu\alpha\beta}  \eta'
 (\sqrt{{2\over3}} cos \theta + \sqrt{{1\over 3}}sin\theta)
\partial_\mu \rho^0_\nu \partial_\alpha \pi^+ \partial_\beta \pi^-.
\end{eqnarray}

 The amplitude of the box diagrams is obtained with $\rho^0 \rightarrow \pi^+ \pi^-$

\begin{equation}\label{eq.13}
\begin{split}
T(2)  = & {1\over g} \frac{8\sqrt{2}}{f^3_\pi}\frac{N_C}{3\pi^2}(\sqrt{{2\over3}} cos \theta +
\sqrt{{1\over 3}}sin\theta) ({1\over2}-{2c\over g}+ {c^2\over g^2}) \\
& m_{\eta'} \epsilon_{ijk} k_{1i} k_{2j} k_{3k} \Big\{\frac{f_{\rho\pi\pi}(q^2_1)}{q^2_1 - m^2_\rho +i\sqrt{q^2_1} \Gamma_\rho (q^2_1)}
 + \frac{f_{\rho\pi\pi}(q^2_2)}{q^2_2 - m^2_\rho +i\sqrt{q^2_2} \Gamma_\rho (q^2_2)}  \\
& - \frac{f_{\rho\pi\pi}(q^2_3)}{q^2_3 - m^2_\rho +i\sqrt{q^2_3} \Gamma_\rho (q^2_3)}
- \frac{f_{\rho\pi\pi}(q^2_4)}{q^2_4 - m^2_\rho +i\sqrt{q^2_4} \Gamma_\rho (q^2_4)}\Big\},
\end{split}
\end{equation}
where \(q_1 = k_1 + k_2,\; q_2 = k_3 + k_4,\; q_3 = k_1 + k_4,\;q_4 = k_2 + k_3 .\)

Adding both the triangle and the box diagrams (Eqs.~(\ref{eq.10},\ref{eq.13})) the total amplitude of the decay
$\eta' \rightarrow\pi^+\pi^-\pi^+\pi^-$ can be obtained.

The $\eta'\rightarrow \pi^+ \pi^- \pi^0 \pi^0$ is another decay mode and have been measured~\cite{15,16}.
The decay mode $\eta'\rightarrow \pi^+ \pi^- \pi^+ \pi^-$ is related to the $\eta' \rightarrow \pi^+ \pi^- \pi^0 \pi^0$
mode by isospin relation. For triangle diagrams, these decays are from the term

\begin{equation}\label{eq.14}
\varepsilon^{\mu\nu\alpha\beta} \partial_\mu \rho^i_\nu \partial_\alpha \rho^i_\beta .
\end{equation}

Ignoring the Lorentz indices we have

\begin{equation}\label{eq.15}
\begin{split}
&\rho^i\rho^i  = 2 \rho^+ \rho^- + \rho^0 \rho^0, \\
&\rho^+ \rho^-  \rightarrow \pi^+ \pi^- 2\pi^0, \\
&\rho^0\rho^0 \rightarrow 2 \pi^+ 2 \pi^-.
\end{split}
\end{equation}

For the box diagrams, the related term can be written as

\begin{eqnarray}\label{eq.16}
\varepsilon^{\mu\nu\alpha\beta} \epsilon_{ijk}\partial_\mu \rho^i_\nu \partial_\alpha\pi^j\partial_\beta\pi^k.
\end{eqnarray}

The isospin structure of Eq.~(\ref{eq.16}) is the same as the $\rho \rho$ (one $\rho$ decays to two pions) of Eq.~(\ref{eq.14}).
Taking the properties of identical particles  (the mass difference between the charged and neutral pions is considered in our uncertainty estimate) this isospin structure predicts

\begin{equation}\label{eq18}
Br(\eta'\rightarrow \pi^+ \pi^- \pi^0 \pi^0) = 2 Br(\eta'\rightarrow \pi^+ \pi^- \pi^+ \pi^-).
\end{equation}


To obtain the branching ratios, we insert the numerical values:

1) Pion decay constant: $\frac{1}{\sqrt{2}}f_\pi = 131.5(1.0)\; MeV$ is taken, where the uncertainty comes from the difference
between the input value of $f_\pi$ and the PDG experimental value~\cite{34}.

2) Universal coupling constant: $g=0.395(8)$ with the uncertainty is assigned to make our prediction of the decay rate $
\Gamma^{th}(\rho \rightarrow e^- e^+) = 6.75(29)\; keV$ to be agreed with the PDG value $\Gamma^{exp}(\rho \rightarrow e^- e^+) = 7.04(6)\;keV$ within 1 $\sigma$~\cite{34}.

3) Weighted average pion mass $M_\pi= (3M_{\pi^+}  +  M_{\pi^0})/4=138.4(1.2)\; MeV$.
To account for isospin breaking effects due to the phase space corrections, we consider the difference between the charged pion mass and weighted average value as uncertainty.


4) Total width of the $\eta'$: $\Gamma_{\eta'}=0.196$ MeV~\cite{34}. To obtain branching ratios we normalize partial widths by this value.

In table (\ref{tab1}), we summarized our predictions for $\eta'\rightarrow \pi^+\pi^-\pi^{+(0)}\pi^{-(0)}$, together with the results of the other theoretical study~\cite{17}, and the experimental values~\cite{16}.

\begin{table}[ht]\label{tab1}
  \centering
  \begin{tabular}{| c | c | c |  }
    \hline
    & $Br(\eta'\rightarrow \pi^+\pi^- \pi^+ \pi^-)\times 10^{-5}$ & $Br(\eta'\rightarrow \pi^+\pi^- \pi^0 \pi^0)\times 10^{-5}$\\
   \hline
   experiment~\cite{16} &  $8.53(0.69)(0.64)$ & $18.2(3.5)(1.8)$ \\
     \hline
   GKW~\cite{17} &$10(3) $ &  $24(7)$\\
  \hline
   This work (triangle diagrams) & $4.2(0.6)$ & $8.3(1.1)$   \\
  \hline
   This work (triangle and box diagrams)& $8.3(1.2)$ & $16.6(2.4)$\\
   \hline
  \end{tabular}
  \caption{Comparison of our predictions with those from the other model, and also the experiment.}
  \label{tab:result}
  \centering
\end{table}

The uncertainty
of our prediction comes from the uncertainties of the input parameters of the theory, $f_\pi$ and $g$, and the weighted average pion mass, which are
combined in quadratic. From Table (\ref{tab1}) we can see that the contribution of the triangle diagrams (Eq.~(\ref{eq.10})) is smaller than the
experiment~\cite{16} and the agreement becomes excellent when the box diagrams are included. 
 In reference~\cite{17} the amplitudes are dominated by the triangle anomaly term.



Of course, there is pentagon diagram for the decay $\eta'\rightarrow \pi^+ \pi^- \pi^+ \pi^-$. It has been shown that the coupling constants
of effective chiral Lagrangian for strong interactions are saturated by meson resonance exchange~\cite{11,35,36}.
There is no $\rho$-resonance in the pentagon diagram and it is believed that its contribution to this decay mode is small.  Thus, calculation of pentagon diagrams is not presented in this work. We are certain that the error made  thereby is well below our uncertainty estimate.


\section{Summary}

In summary, an effective chiral theory has been applied to study the two decay modes of $\eta'\rightarrow \pi^+ \pi^- \pi^+ \pi^-$ and
$\eta'\rightarrow \pi^+ \pi^- \pi^0 \pi^0$. In this work we have evaluated  the triangle and box anomalous diagrams. We have also shown that the
contribution of box diagrams is important in these processes.
Theoretical predictions are: $Br(\eta' \rightarrow \pi^+ \pi^- \pi^+ \pi^-)= {1 \over 2} Br(\eta' \rightarrow \pi^+ \pi^- \pi^0
\pi^0)= (8.3 \pm 1.2) \times 10^{-5}$, which are in good agreement with experimental data: $Br(\eta' \rightarrow \pi^+ \pi^- \pi^+
\pi^-)=[8.53\pm0.69(stat.)\pm0.64(syst.)] \times 10^{-5}$ and $Br(\eta' \rightarrow \pi^+ \pi^- \pi^0 \pi^0)=[1.82\pm0.35(stat.)
\pm0.18(syst.)]\times 10^{-4}$.


\section*{References}

\bibliography{mybibfile}

\end{document}